\documentclass[12pt,preprint]{aastex}
\newcommand{\rxte}{{\em RXTE\ }}
\newcommand{\rxteno}{{\em RXTE}}
\newcommand{\exosat}{{\em EXOSAT\ }}
\newcommand{\exosatno}{{\em EXOSAT}}
\newcommand{\integral}{{\em INTEGRAL\ }}
\newcommand{\integralno}{{\em INTEGRAL}}
\newcommand{\swift}{{\em Swift\ }}
\newcommand{\swiftno}{{\em Swift}}

\begin{document}
\title{Outbursts Large and Small from EXO 2030+375}
\author{Colleen A. Wilson, Mark. H. Finger}
\affil{VP62, National Space Science and Technology Center, 320 Sparkman Drive, Huntsville, AL 35805}
\email{colleen.wilson@nasa.gov}
\author{Ascensi{\'o}n Camero Arranz}
\affil{GACE/ICMUV, Universidad de Valencia, P.O. Box 20085, 46071 Valencia,
Spain}
\begin{abstract}

During the summer of 2006, the accreting X-ray pulsar EXO 2030+375 underwent
its first giant outburst since its discovery in 1985. Our 
observations include the first ever of the rise of a giant outburst of  EXO 
2030+375. EXO 2030+375 was monitored daily with the 
{\em Rossi X-ray Timing Explorer (RXTE)} from 2006 June through 2007 May. During the 
giant outburst, we discovered evidence for a cyclotron
feature at $\sim 11$ keV. This feature was confidently detected for about 90
days, during the brighter portion of the outburst. Daily observations of the 
next five EXO 2030+375 orbits detected pulsations at all orbital phases and
normal outbursts shifted to a later orbital phase than before the giant 
outburst. An accretion disk appears to be present in both the normal and giant
outbursts, suggesting that the long-term behavior is a product of the state of
the Be star disk and the accretion disk. Here we will present flux and frequency histories from our detailed 
\rxte observations of the giant outburst and the normal outbursts that
surrounded it. A new orbital analysis is presented that includes observations from 1991 
through 2007 August. 

\end{abstract}
\section{Introduction}
Be/X-ray binaries are the most common type of accreting X-ray pulsar systems
observed.
They consist of a pulsar and a Be (or Oe) star, a main sequence star of
spectral type B (or O) that shows Balmer emission lines 
\citep[See e.g.,][for a  review.]{Porter03} The line emission is believed to be 
associated with circumstellar material shed by the Be star into its equatorial 
plane. The exact nature of the mass loss process is unknown, but it is thought to be related to
the rapid rotation of the Be star, typically near 70\% of the critical break-up
velocity \citep{Porter96}. The equatorial material forms a slow, dense outflow,
which is generally believed to fuel the X-ray outbursts. Near the Be star, the
equatorial outflow probably forms a quasi-Keplerian disk
\citep{Quirrenbach97,Hanuschik96}.

X-ray outbursts are produced when the pulsar interacts with the Be star's disk.
Be/X-ray binaries typically show two types of outburst behavior: (a) giant 
outbursts (or type II), characterized by high luminosities and high spin-up 
rates  (i.e., a significant increase in pulse frequency) and (b) normal 
outbursts (or type I), characterized  by lower luminosities, low spin-up rates
(if any), and repeated occurrence at the orbital period \citep{Stella86,Bildsten97}. As a
population Be/X-ray binaries show a  correlation between their spin and orbital
periods \citep{Corbet86,Waters89}.

For isolated Be stars, variations in the infrared bands (J,H,K) are believed to be good
indicators of the size of the Be star's disk. However, when the Be star is in a
binary system with a neutron star, the Be disk is truncated at a resonance 
radius by tidal forces from the orbit of the neutron star \citep{Okazaki01}. In
these systems, since the disk cannot easily change size because of the truncation 
radius, changes in mass loss from the Be star produce changes in the disk 
density, which can even become optically thick at infrared wavelengths 
\citep[see, e.g.,][]{Neg01a,Mir01}.  

EXO 2030+375 is a 42-s transient accreting X-ray pulsar discovered with {\em
EXOSAT} during a giant outburst in 1985 \citep{Parmar89a}. In this system, the
pulsar orbits a B0 Ve star \citep{Motch87, Janot88, Coe88} every 46 days
\citep{Wilson05}. A normal outburst has been detected for nearly every periastron
passage since 1991 \citep{Wilson02, Wilson05}. For these normal outbursts, the
outburst intensity and the global spin-up rate appeared to be tied to
the K-band intensity of the Be star. From 1992-1994 EXO 2030+375's outbursts
were bright and the pulsar was spinning-up. In 1994, shortly after a drop in the
K-band intensity of the Be disk, the X-ray intensity abruptly dropped and the
global trend changed to spin-down, indicating that the density of the Be disk
had dropped and less material was available for accretion. The pulsar continued
with faint X-ray outbursts and a global spin down trend until 2002, when again
the K-band intensity increased to the 1992-1994 level followed by brighter 
X-ray outbursts and a transition to global spin-up \citep{Wilson05}. The 
outbursts continued to brighten and show spin-up until 2006 June when EXO 
2030+375 underwent its first giant outburst since its discovery in 1985 
\citep{Corbet06,Krimm06,McCollough06,Wilson06}. In this paper we present a
timing analysis including twelve normal outbursts leading up to the giant
outburst, the giant outburst, and seven normal outbursts after the giant
outburst combined with previously published observations.

Accreting X-ray pulsars have strong surface magnetic fields of $\sim 10^{12}$ G. A
direct measurement of this field strength is provided by the energies of
cyclotron resonance scattering features in their X-ray spectra. The
magnetic field strength and the resonance energy are related as $E_{\rm cyc} = 11.6
B_{12} (1+z)^{-1}$, where $B_{12}$ is the magnetic field strength in units of
$10^{12}$ G and $z$ is the combined gravitational plus bulk motion Doppler shift 
\cite[and references therein]{Nakajima06}. The range of previously measured cyclotron
features is from about 11 keV in 4U0115+63 \cite{Nakajima06} to  about 50 keV in
A0535+262 \citep{Terada06}. Previously a tentative cyclotron at 36 keV feature was
reported in a 1996 normal outburst of EXO 2030+375 \citep{Reig99}; however, this 
feature has not been seen in other any other observations. Recent observations 
of EXO 2030+375 during its giant outburst have resulted in reports of a 
cyclotron feature near 10 keV \citep{Wilson06, Klochkov07}. In this paper we
present detailed spectral fits to \rxte observations from the giant 
outburst, demonstrating that the cyclotron feature was consistently detected 
over an extended period of time. 

%EXO 2030+375 is a 42-s transient accreting X-ray pulsar with a Be star companion
%\citep{Coe88, Janot88} that was discovered during a giant outburst with 
%{\em EXOSAT} \citep{Parmar89a,Parmar89b}. The pulsar orbits its companion in a
%46-day eccentric orbit and showed outbursts near every observed periastron
%passage since 1991 \citep{Wilson02,Wilson05}.

%Be/X-ray binaries such as EXO 2030+375 show two types of outbursts: (1) Normal
%or Type I outbursts that are periodic with the orbital period, resulting from
%interaction between the pulsar and the Be star's circumstellar disk and (2)
%Giant or Type II outbursts that exhibit much larger fluxes and spin up
%rates\citep{Stella86,Bildsten97}.

\section{Observations and Analysis}

\begin{figure}
\plotone{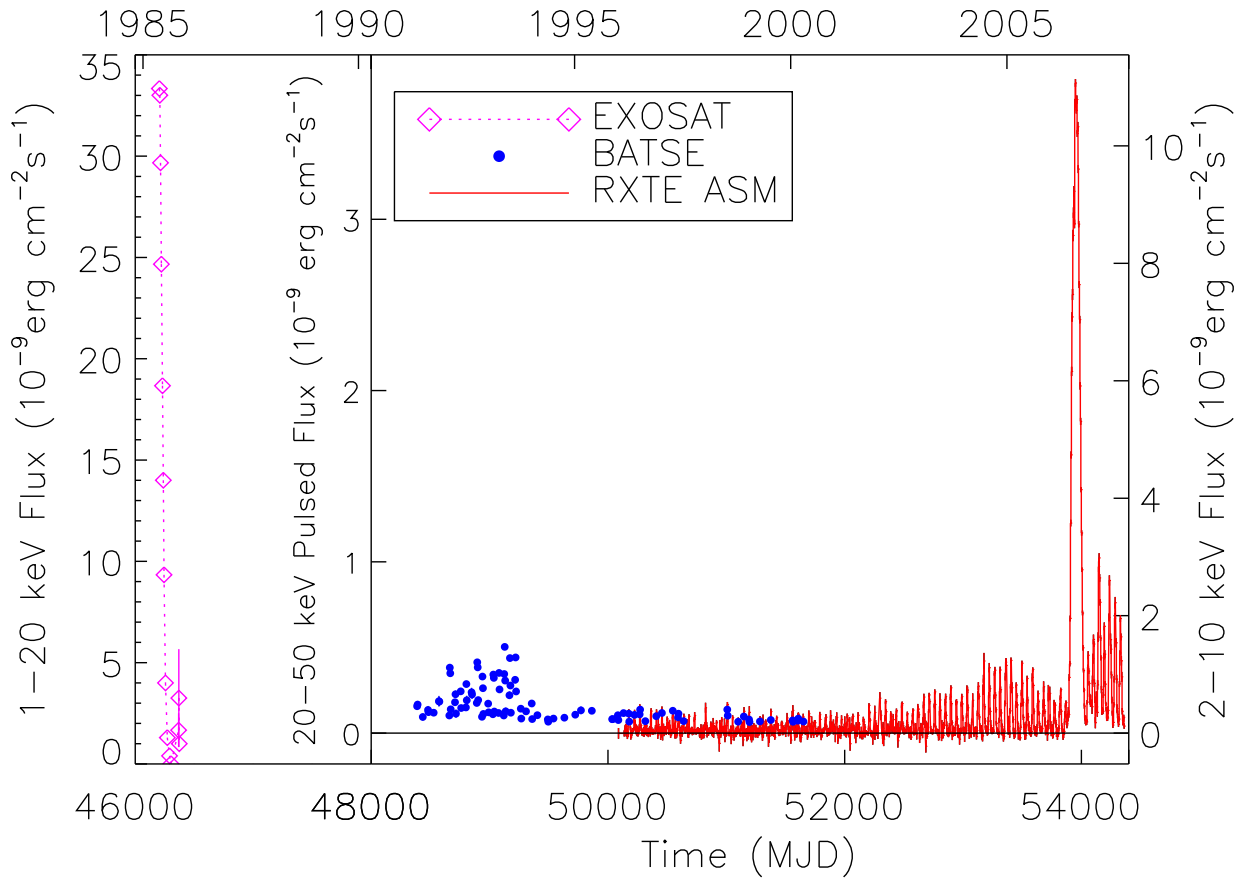}
\caption{Long-term Flux history for EXO 2030+375. The pink diamonds and left
y-axis denote \exosat 1-20 keV total flux measurements from \citet{Parmar89a}.
Blue points and the y-axis near the center denote 20-50 keV pulsed fluxes
measured with BATSE. Typically 1-4 of these BATSE points correspond to a normal outburst.
The red histogram and right y-axis denote \rxte ASM 2-10 keV total flux 
measurements averaged over 4 days. Each spike corresponds to a normal outburst 
and the large peak is the 2006 giant outburst. 
\label{fig:ltflx}}
\end{figure}

Figure~\ref{fig:ltflx} shows the long-term flux history for EXO 2030+375 since
its discovery as measured with \exosat, BATSE, and \rxte ASM. This Figure is provided here to aid the reader
in placing the observations described below in context.

\subsection{\rxte Observations}

The 2006 giant outburst and five normal outbursts that followed from EXO 2030+375
were observed daily from 2006 June 22 through 2007 May 25 (MJD 53,908-54,245) with
the {\em Rossi X-ray Timing Explorer (\rxteno)} Proportional Counter Array 
\citep[PCA]{Jahoda06} and the High-Energy X-ray Timing Experiment 
\citep[HEXTE]{Rothschild98}. Also included in this paper are analyses of previously
unpublished \rxte observations of eight normal outbursts in the year prior to the 
giant outburst (MJD 53,540-53,868) and \rxte observations of outbursts in 2007
June (MJD 54,276-54,283) and 2007 August (MJD 54,322-54,329). For these outbursts,
the typical 1-5 ks observations spanned about six days near the peak of the 
normal outburst. In this paper, timing analysis was performed using the Standard 1 
PCA data, which has 0.125 s time resolution and no spectral resolution. Phase 
averaged spectral analysis was performed using the Standard 2 PCA data with 129 
channel energy resolution and 16-s time resolution and the science data mode
E\_8us\_256\_DX1F HEXTE data for cluster B only, since cluster A was not rocking and
was not on-source during the early and late parts of the giant outburst.

\subsection{\integral Observations}
{\em International Gamma Ray Astrophysics Laboratory} (\integralno) ISGRI
\citep{Ubertini03} public data from revolutions 18-22 (MJD 52,615-52,627), 67
(MJD 52,661-52,662), 80 (MJD 52,800-52,802), 159-160 (MJD 53,036-53,041), and 
185-193 (MJD 53,126-53,133) was also included in our timing analysis. Our software,
described in detail for the revolutions 18-22 observations in \citet{Camero05}, 
collected good events in the 20-60 keV band using the pixel information function 
when the source was in the partially or fully coded field-of view of ISGRI. The 
good events were then epoch folded using a simple phase model based on spin 
frequency measurements with \rxte and \integral. To avoid binning effects, each 
pulse profile was fit with a Fourier series of harmonic coefficients. A correction
described in \citet{Camero05} was applied to account for aperiodic noise due to 
pulse profile variations and due to nearby noisy sources. A template profile was 
estimated from the average profile for revolutions 18-22. To generate phase offsets
from the model, we then cross-correlated the individual profiles with the template
profile. The new phases for each outburst were then fit with a linear or
quadratic phase model and the process was repeated, creating new folded
profiles, new harmonic coefficients, and new phase offsets. The pulse profiles
were then combined over time to improve statistics and to allow the phase
measurements to constrain spin-up during each outburst. Pulse phase measurements
with \integral from five normal outbursts of EXO 2030+375 are used in this paper.

\subsection{Timing Analysis}
For each \rxte PCA observation, we generated a light curve file from Standard 1
data using FTOOLS v6.2. We corrected each light curve to one average PCU using
the FTOOL {\it correctlc}. Using {\it faxbary} we corrected the times on
each bin to the solar system barycenter. Then we corrected the times for
the pulsar's orbit using the orbital parameters from \citet{Wilson05}. Lastly, 
we fit a pulse profile model consisting of a 6th order Fourier expansion in 
the pulse phase model plus a constant background term. For outbursts before and
after the giant outburst where we did not have daily coverage, the phase
model initially consisted of a constant frequency and was iteratively improved
to a quadratic phase model for each outburst. For the daily 
measurements, the model was a quadratic spline with each $\sim 5$-day interval 
having an independent spin-up rate. To compensate for aperiodic noise due to pulse
profile variations within each \rxte observation, we first subtracted the pulse
profile model from the light curve and then computed a power spectrum. Within 
each power spectrum we computed the average noise power around each harmonic 
$\bar P_n$ in the frequency range $[(n-1/2)\nu,(n+1/2)\nu]$ where $n$ is the 
harmonic number and $\nu$ is the model frequency for that observation. The 
errors on the pulse profile harmonic coefficients were then inflated by 
$(\bar P_n/2)^{(1/2)}$ where 2 is the assumed Poisson level. Phase offsets were 
computed by cross-correlating the individual profiles with a template profile.

Examination of the measured spin-up rates within the contiguous daily
observations of the giant outburst and five normal outbursts showed a periodic 
dip in the spin-up rate, just before periastron, indicating that the orbital 
solution needed improvement. To improve the orbital solution, we combined our 
new phase measurements with previously published phase measurements from the 
Burst and Transient Source Experiment (BATSE)
\citep[51 outbursts: 1991-2000 (MJD 48,385-51,657)]{Wilson02} \rxte 
\citep[4 outbursts; 1996 (MJD 50,266-50,275), 1998 (MJD 50,821-50,828), 2002
(MJD 52,432-52,443), and 2003 (MJD 52,894-52,899)]{Wilson02, Wilson05}, and 
\integral \citep[1 outburst; 2002]{Camero05}. The phases were fit with a global
orbital model plus a quadratic spline. Prior to the giant outburst and for
outbursts from 2007 June through September, the quadratic spline had an 
independent frequency derivative for each outburst. Phase ``slips", jumps in the
pulse cycle count, were fit in large gaps between outbursts where no data were 
available to constrain the phase model. The large spin-up rates during the giant
outburst overwhelmed the orbital effects, so it was excluded from orbital 
analysis. 

We found, however, that our initial fits were much poorer than expected,
with reduced $\chi^2$ values of about 2.0. Neither the phase residuals nor the
phase errors showed any dependence on intensity or orbital phase, suggesting
that the phase errors were simply too small across the board, indicating that
the phase errors did not sufficiently reflect the effects of pulse profile
variations from observation to observation. We added a systematic error of 
$2.4 \times 10^{-3}$ cycles in quadrature to all of the phase measurements 
included in our fits. In addition we eliminated six BATSE points that were large
outliers despite the already large BATSE error bars. 

Table~\ref{tab:orb} lists the orbital parameters resulting from our fits: the orbital period $P_{\rm
orb}$, the epoch of periastron passage $T_{\rm peri}$, the projected semi-major
axis $x = a_x \sin i$, the eccentricity $e$, and the periapse angle $\omega$. The
largest differences between our new orbital fit (Fit 1) and published results
\citep{Wilson05} are in $e$ (2.9$\sigma$ smaller) and $a_x \sin i$ (2.5$\sigma$
larger). Next we searched for time dependent variations in the orbital
parameters, fitting the orbital period derivative $\dot P_{\rm orb}$, the 
derivative of the periapse angle $\dot \omega$, and the derivative of the 
projected semi-major axis $\dot x$. First we varied each parameter separately,
holding the other two fixed at zero, while allowing all other fit parameters to
vary. Varying $\dot \omega$ produced the most significant result (Fit 2), with
an F-test significance of 3.4$\sigma$. $\dot P_{\rm orb}$ and $\dot x$
were less significant with F-test values of 1.7$\sigma$ and 3.0$\sigma$,
respectively. Next we tried varying both $\dot \omega$ and $\dot x$ (Fit 3) while
holding $\dot P_{\rm orb}$ fixed at zero and allowing all other parameters
to vary. This fit had an F-test significance of 3.6$\sigma$ for two parameters
and 2.0$\sigma$ for the addition of $\dot x$. Lastly we tried varying all three
parameters (Fit 4). This fit had an F-test significance of 3.7$\sigma$ for the
addition of three parameters and 2.0$\sigma$ for just the addition of $\dot
P_{\rm orb}$.

\begin{deluxetable}{lllll}
\tablecaption{EXO 2030+375 Orbital Fits}
\tablewidth{0pt}
\tablehead{\colhead{Parameter} & 
\colhead{Fit 1} & \colhead{Fit 2} & \colhead{Fit 3} & \colhead{Fit 4}}
\startdata
$P_{\rm orb}$ (days) &  $46.0205 \pm 0.0002$ & $46.0213 \pm 0.0003$ & 
 $46.0211 \pm 0.0003$ & $46.0207 \pm 0.0004$ \\
$T_{\rm peri}$ &  $54044.73 \pm 0.01$ & $52756.17 \pm 0.01$
 & $52802.20 \pm 0.01$  & $53308.45 \pm 0.02$ \\ 
$x$ (lt-s) &  $244 \pm 2$ & $246 \pm 2$ & $248 \pm 2$ & $248 \pm 2$\\
$e$ &  $0.412 \pm 0.001$ & $0.410 \pm 0.001$ & $0.409 \pm 0.001$ & 
 $0.410 \pm 0.01$\\
$\omega$ (deg) &  $211.3 \pm 0.3$ & $211.9 \pm 0.4$ & 
 $212.0 \pm 0.4$ & $212.6 \pm 0.4$\\
$\dot P_{\rm orb}$ (days/day) & \nodata & \nodata & \nodata & 
 $(-4 \pm 2) \times 10^{-7}$\\
$\dot \omega$ (deg/year)  & \nodata & $0.18 \pm 0.05$ & $0.15 \pm 0.05$ & 
 $0.17 \pm 0.06$ \\
$\dot x$ (lt-s/year)  & \nodata & \nodata & $0.2 \pm 0.1$ & $0.2 \pm 0.1$\\
$\chi^2/$dof &  689.8/656 & 677.5/655 & 673.4/654 & 669.4/653 \\
\enddata
\label{tab:orb}
\end{deluxetable}

\begin{figure}
\plotone{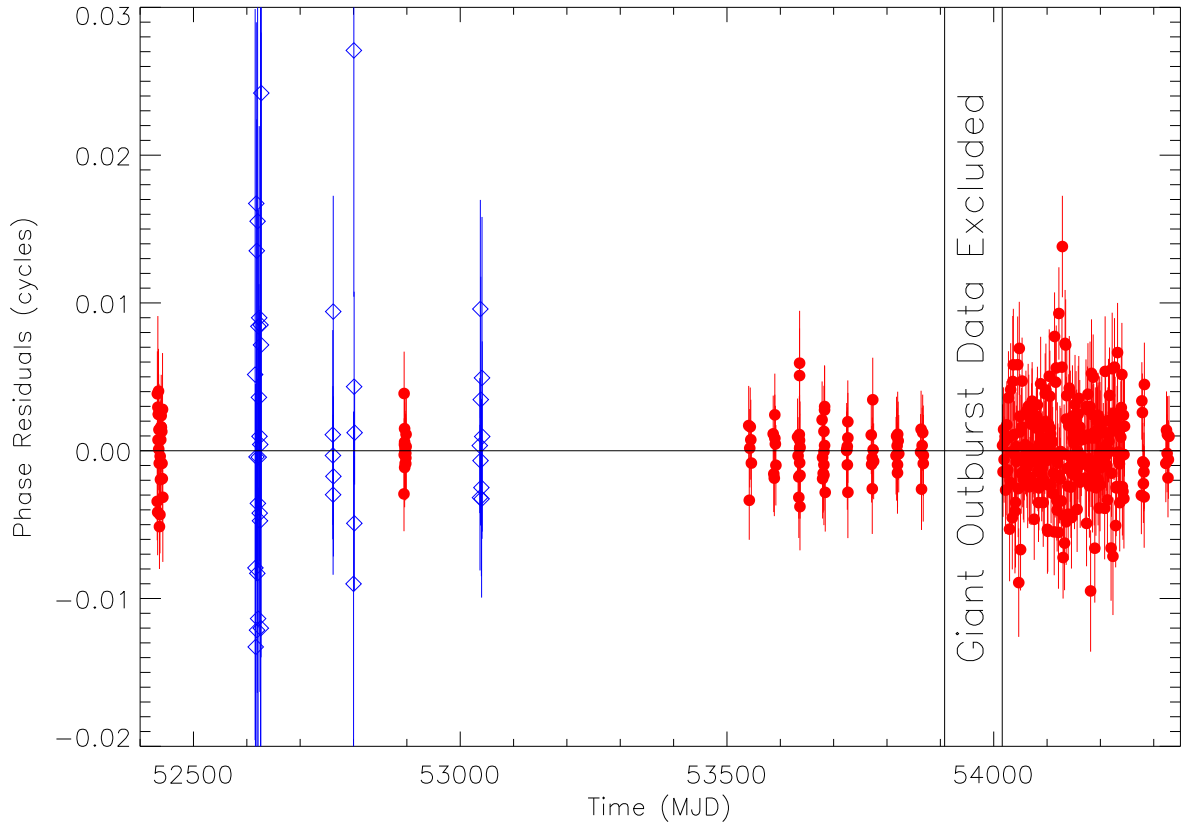}
\caption{Phase residuals for all of the measurements after May 2000 included in
the fitting. Red filled circles denote \rxte measurements and blue open diamonds
denote \integral measurements. Phase residuals from BATSE and \rxte before 
May 2000 are shown in \citet{Wilson02}. The differences between this orbital 
fit and that one are too small to be visually apparent. \label{fig:postbatseres}}
\end{figure}

\begin{figure}
\plotone{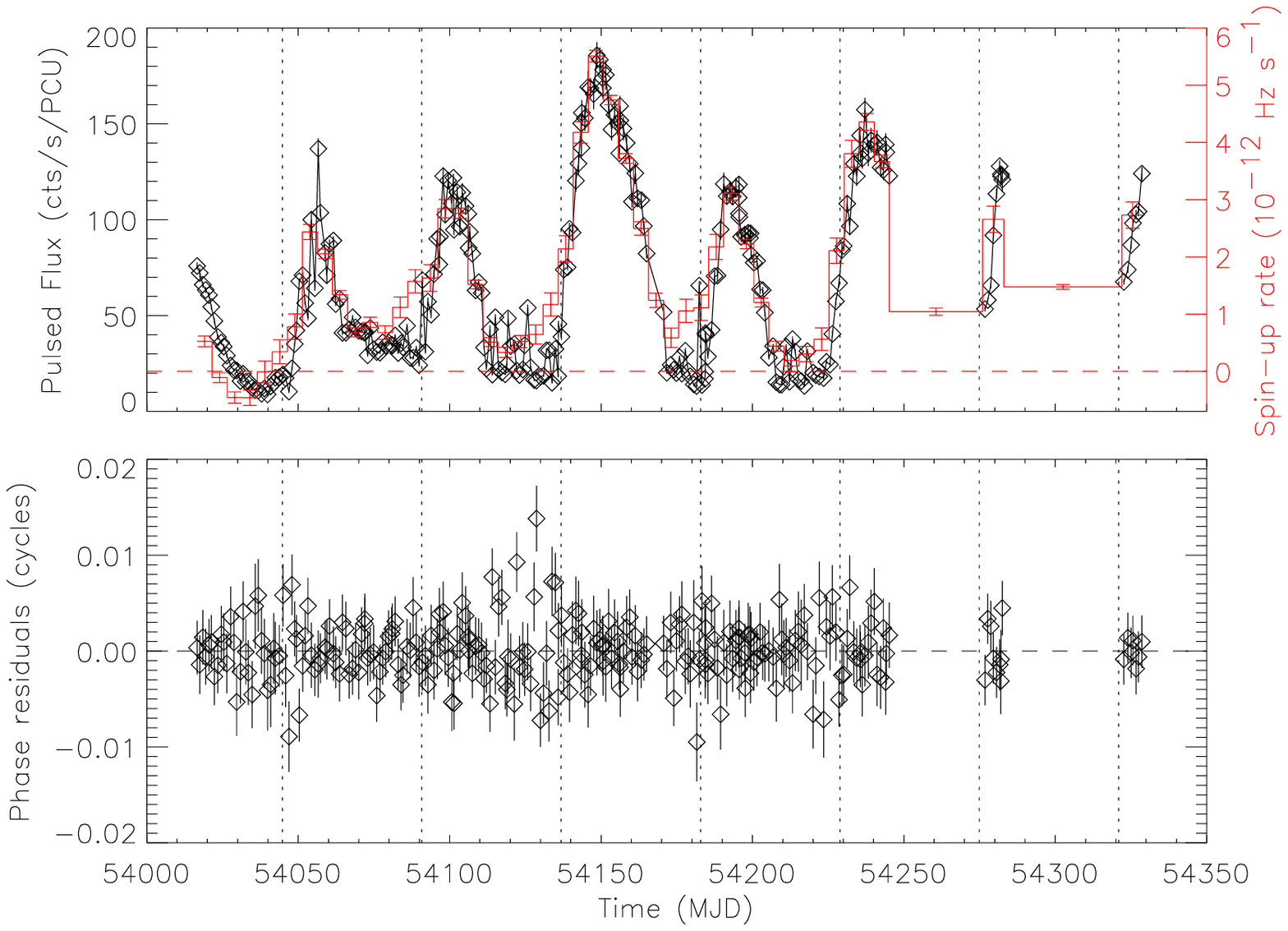}
\caption{A close look at timing results after the giant outburst. {\it Top:} 
2-60 keV rms pulsed flux measured with the \rxte PCA (left y-axis, black diamonds) 
overlaid with the estimated model frequency derivatives (right y-axis, red 
histogram). The red dashed line denotes a frequency derivative equal to zero. 
{\it Bottom:} Phase residuals after the giant outburst. \label{fig:postgiantres}}
\end{figure}

Figure~\ref{fig:postbatseres} shows the phase residuals for all phase
measurements after BATSE was de-orbited in May 2000. Phase residuals prior to
this date are shown in \citet{Wilson02}.
Figure~\ref{fig:postgiantres} zooms in to show the 2-60 keV rms pulsed flux 
overlaid with the frequency derivatives estimated from the orbit fitting and the phase
residuals. From this figure, we note that the spin-up rate and pulsed flux are
clearly correlated across all orbital phases. At the end of the giant outburst,
the pulsar spins down briefly. The minimum in the frequency derivative precedes
the pulsed flux minimum slightly. Although there are clearly defined
outbursts for each orbit, the pulsed flux never drops to zero between outbursts and we see
considerable flaring activity in the intra-outburst region. 

\begin{figure}
\epsscale{0.8}
\plotone{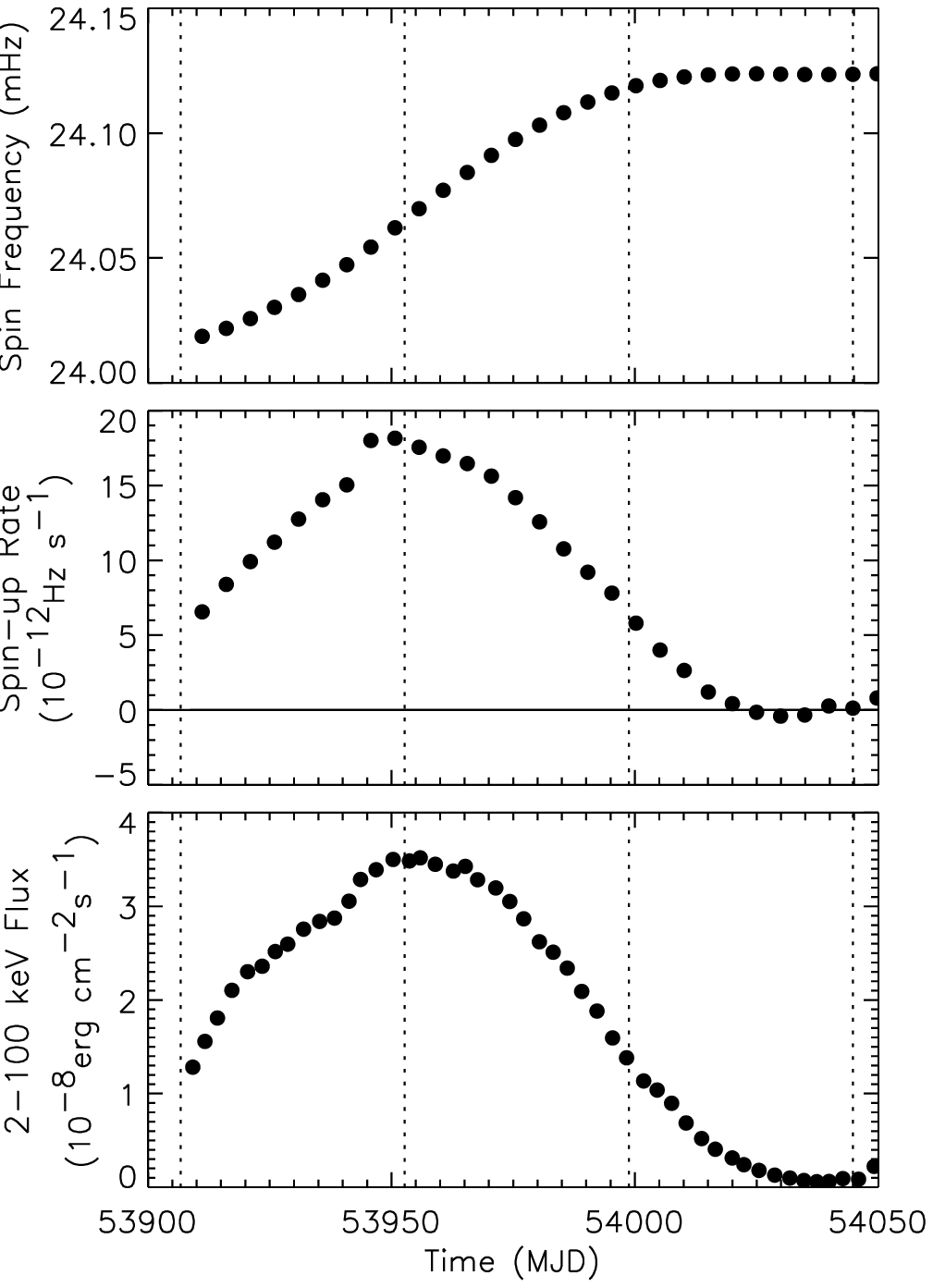}
\caption{Measured spin-frequency, spin-up rate, and 2-100 keV flux for the 2006
giant outburst from \rxte 
data. Each frequency and spin-up rate point corresponds to the average 
barycentered, orbit corrected (using Fit 3), frequency and frequency derivative
for a 5-day interval of \rxte PCA observations. The bottom panel shows 3-day
averages of the total 2-100 keV flux measured with joint spectral fits to \rxte
PCA and HEXTE data.\label{fig:giantfreqfdot}}
\end{figure}

Using the orbit from Fit 3 in Table~\ref{tab:orb}, we fit the phase measurements
for the giant outburst with a new quadratic spline model in which each 5-day 
interval had an independent frequency derivative. 
Figure~\ref{fig:giantfreqfdot} shows the spin frequency, the spin-up rate, and
the 2-100 keV flux determined for the giant outburst. The giant outburst reached a peak
spin-up rate of $(1.815 \pm 0.006) \times 10^{-11}$ Hz s$^{-1}$ for the 5-day
interval 2006 August 1-6 (MJD 53948-53953). The 2-100 keV flux also peaked in
this interval at $(3.59 \pm 0.07) \times 10^{-8}$ erg cm$^{-2}$ s$^{-1}$.

\subsection{X-ray Spectral Analysis}
Data analysis was performed using FTOOLS v6.1.2
\citep{Blackburn95}\footnote{\url{http://heasarc.gsfc.nasa.gov/ftools/}}.
For each \rxte PCA observation, we created a background spectrum using the
bright source model with the FTOOL {\it pcabackest} and extracted source and 
background spectra using the FTOOL {\it saextract}. In addition we corrected for
deadtime following the recipe
\footnote{\url{http://heasarc.gsfc.nasa.gov/docs/xte/recipes/cook\_book.html}}. 
A 0.5\% systematic error was included. 
For HEXTE cluster B data, detectors 0,1, and 3, we separated the data into 
source and background using the FTOOL {\it hxtback} and extracted source and 
background spectra using {\it seextrct}. HEXTE spectra were
corrected for deadtime using {\it hxtdead}. An analysis that includes HEXTE
cluster A data is described in \citet{Camero07}. The results were consistent
with ours. Unfortunately cluster A was not on-source during the early and late
parts of the giant outburst, so the data set is more limited. Six observations (obsids
91089-01-01-04, 91089-01-02-10, 91890-01-03-10, 91089-01-04-02, 91089-01-12-01, 
91089-01-15-04) were excluded from our analysis because they had less than 20 seconds
of HEXTE cluster B background exposure, i.e. HEXTE cluster B was not rocking. Only the last two of these observations were 
during the giant outburst.

From a joint fit to \rxte PCA and HEXTE data from August 15, 2006 (MJD 53962),
we found evidence for a cyclotron scattering feature near 10 keV
\citep{Wilson06}. The model
used was an absorbed power-law (PHABS*POWERLAW) with a high-energy cut-off
(HIGHECUT), an iron line (GAUSSIAN), and a
Gaussian cyclotron absorption line (GABS). The power-law had a photon index of
1.53(2), with a cutoff energy at 12.4(4) keV, a folding energy of 27.4(5) keV.
The cyclotron energy was 10.1(2) keV, with a Gaussian width of 3.3(2) keV and a
peak depth of 1.1(1). This feature was significant at a 7.5 sigma level. Other 
continuum models, e.g., a Bremsstrahlung model,also showed evidence for this feature. 

\begin{figure}
\epsscale{1.0}
\plotone{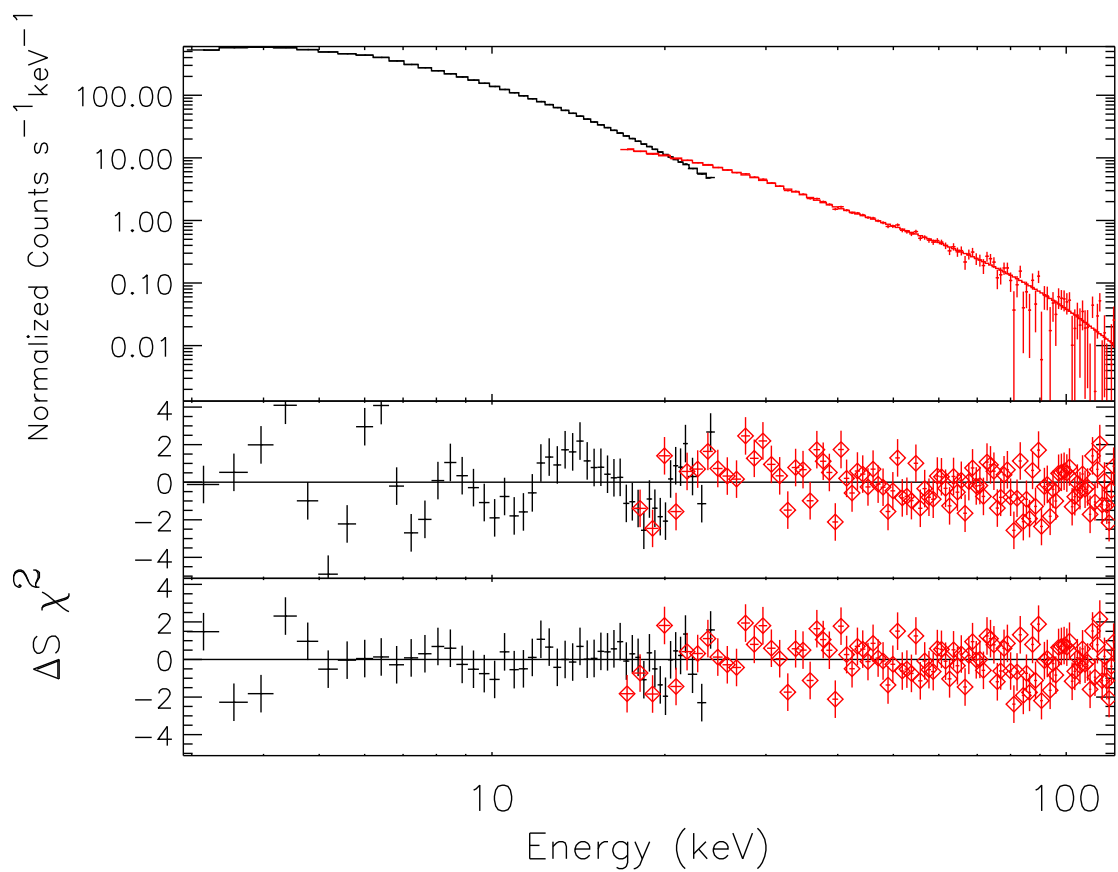}
\caption{Energy spectra from 2007 August 7, near the peak of the
outburst. In the top panel, the data were fit with a model consisting of an 
absorbed power law with a high energy cutoff with a Gaussian absorption feature
representing the cyclotron feature and a Gaussian iron line. The center panel 
shows the residuals in units of sigmas for a fit that excluded the cyclotron
feature. The bottom panel shows the residuals in units of sigmas with the
cyclotron feature included in the model. 
\label{fig:comparespec}}
\end{figure}

\begin{figure}
\epsscale{0.8}
\plotone{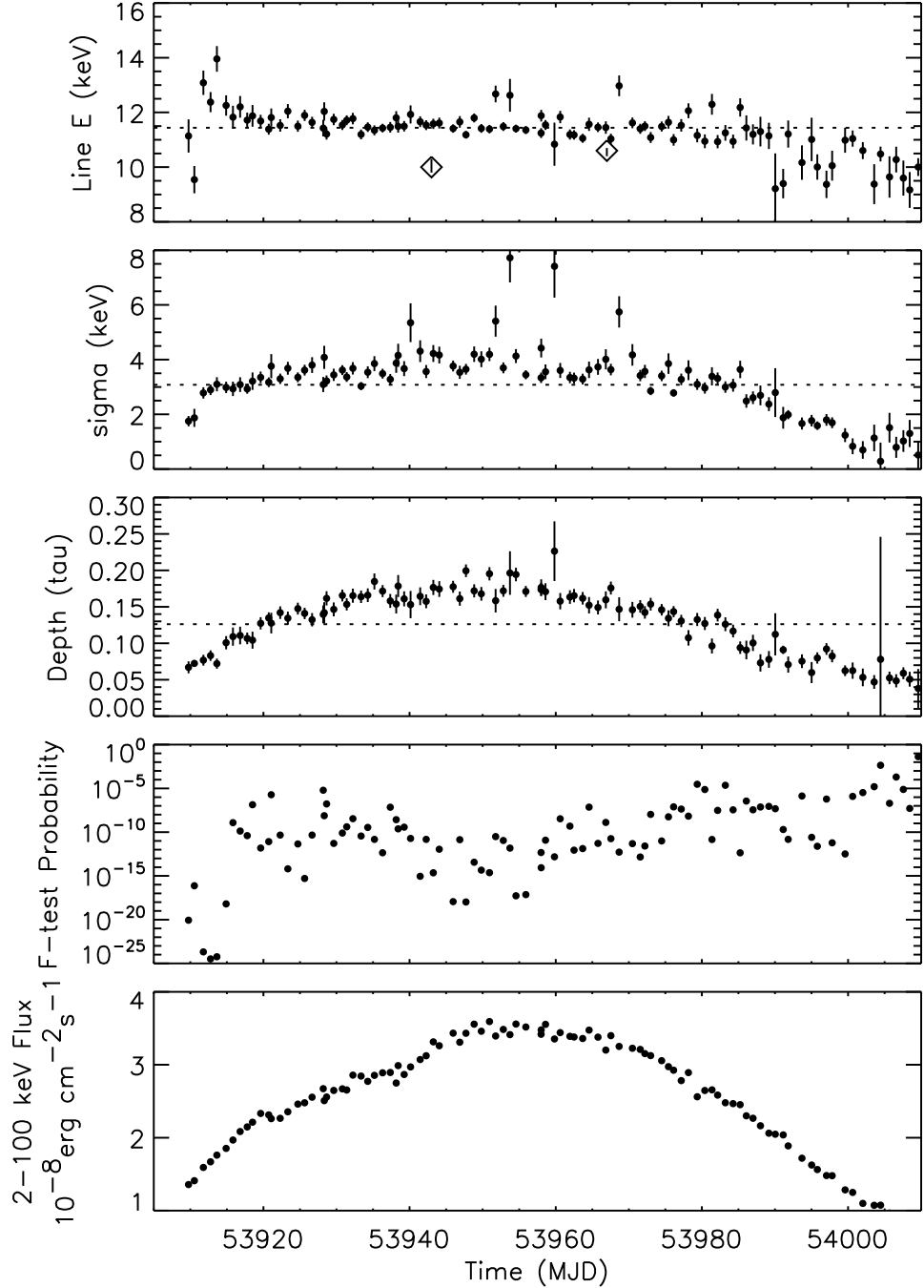}
\caption{From top to bottom: EX0 2030+375 Cyclotron line energy (keV), 
width (keV), and optical depth defined in Equations 6 and 7 in \citet{Coburn02}.
In the top panel, diamond symbols denote the line energy reported in 
\citet{Klochkov07}. Dotted lines show the average value for each parameter. The second from the
bottom panel shows the F-test value in sigmas for including the cyclotron
feature and the bottom panel shows the 2-100 keV flux.}
\label{fig:cycl}
\end{figure}

\begin{figure}
\plotone{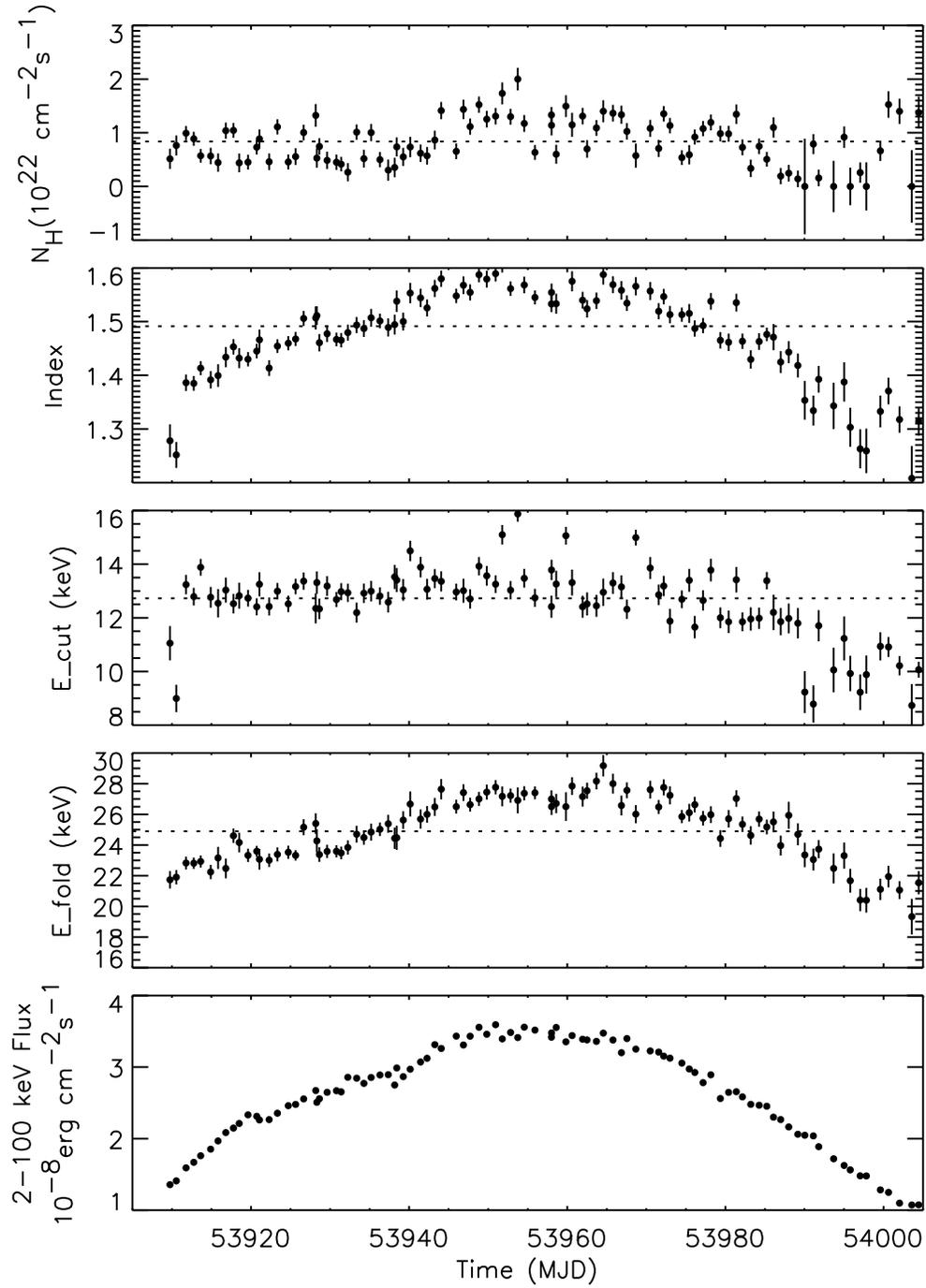}
\caption{Additional model parameters, from top to bottom: the absorption $N_{\rm
H}$, power-law photon index, Cut-off energy $E_{\rm cut}$, e-folding energy
$E_{\rm fold}$, and the 2-100 keV flux.}
\label{fig:otherpar}
\end{figure}

\begin{figure}
\plotone{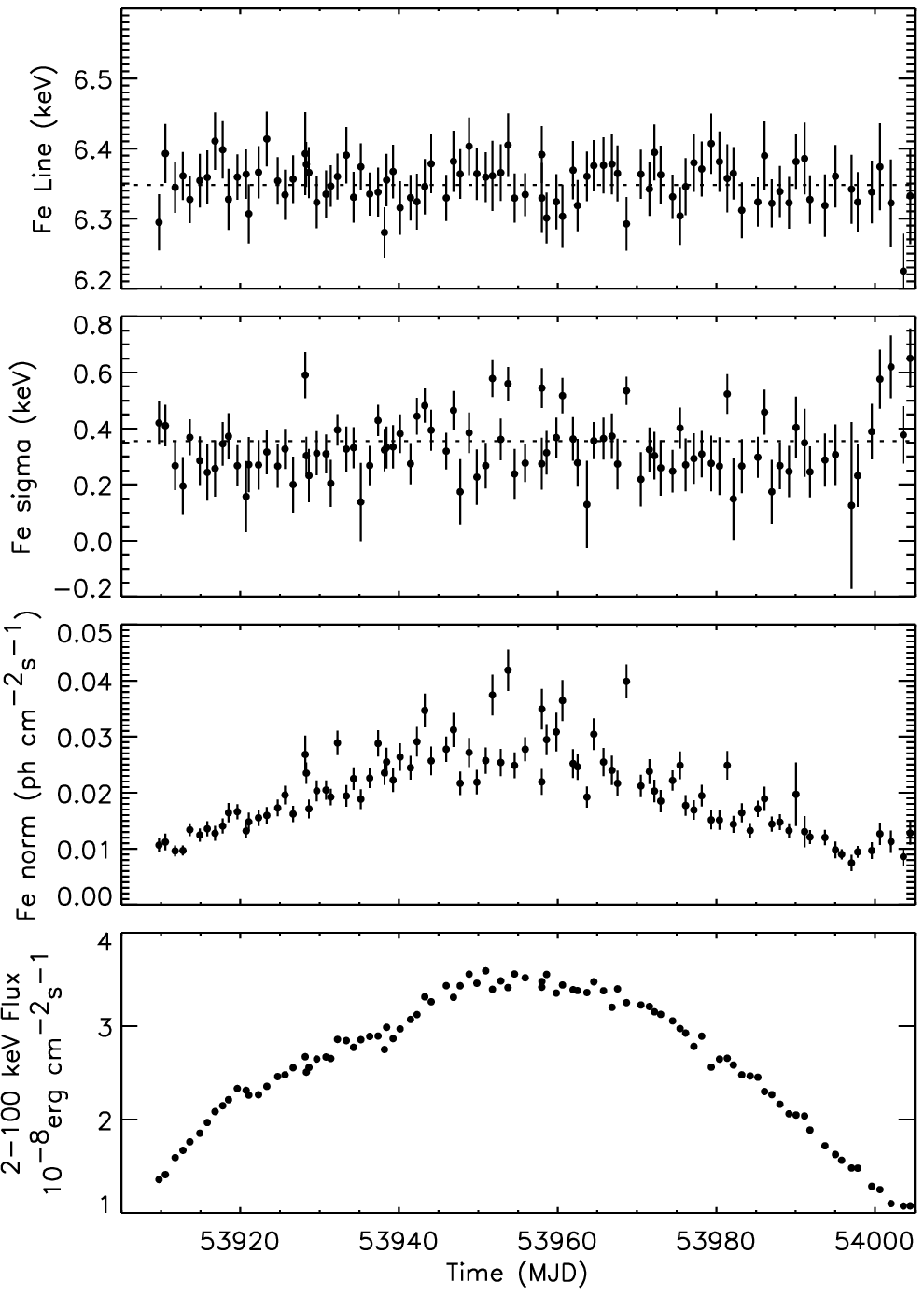}
\caption{Gaussian Iron line model parameters, from top to bottom: the line
energy (keV), line width (keV), the line normalization (photons cm$^{-2}$
s$^{-1}$), and the 2-100 keV Flux.}
\label{fig:Fe}
\end{figure}

%We then fit the PCA and HEXTE data from each observation with an absorbed 
%power-law with a Gaussian iron line and a high energy cutoff. We then added a 
%Gaussian absorption feature (GAUSSABS) to represent a cyclotron line. We 
%discovered a problem with the GABS model in XSPEC that affected the line depth,
%so we wrote our own local model using the parameterization of \citet{Coburn02}.
%Figure~\ref{fig:spec} shows the cyclotron line parameters, the f-test 
%probability of the cyclotron feature,and the 2-100 keV flux. A significant cyclotron feature was only detected during
%the giant outburst, at 2-100 keV fluxes larger than $10^{-8}$ erg cm$^{-2}$
%s$^{-1}$. Of the 102 available observations during the giant outburst, only
%eight are excluded from this plot. Six had less than 20 seconds of HEXTE
%background data and two near the end of the outburst had very large error bars
%for $E_{\rm cyc}$. 
Using Xspec v12.3, we first fit the PCA and HEXTE data from each observation with an absorbed
power-law with a high energy cutoff plus a Gaussian iron line
(PHABS (POWERLAW+GAUSSIAN) HIGHECUT). Based upon our previous detection of a
cyclotron feature at 10 keV \citep{Wilson06}, we included a Gaussian absorption
feature (GAUSSABS) to represent the cyclotron line. We discovered a problem with
the built-in XSPEC model GABS. This model calls the Gaussian line model and then
exponentiates it. However, the Gaussian line model returns the integral of the
Gaussian across the input bins rather than the value of the Gaussian, resulting in
GABS being incorrectly integrated. We wrote a local model GAUSSABS using
the parameterization of \citet{Coburn02}. Figure~\ref{fig:comparespec} shows the 
spectrum and residuals with and without the Gaussian absorption feature for an 
observation on 2006 August 7 (MJD 53954.5) near the peak of the outburst.
Dips are visible at 10 and 20 keV. Figure~\ref{fig:cycl} shows the
cyclotron feature parameters: the line energy $E$, the line width $\sigma$, and
the optical depth $\tau$ defined in Equations 6 and 7 in \citet{Coburn02}.
Dotted lines across each panel show the average value for that parameter 
($\bar E = 11.44 \pm 0.02$ keV, $\bar \sigma = 3.08 \pm 0.02$ keV, and $\bar
\tau = 0.1263 \pm 0.0009$).
The second from the bottom panel in Figure~\ref{fig:cycl} shows the F-test probability
for including the cyclotron feature. The bottom panel shows the
2-100 keV flux for each observation. Figure~\ref{fig:otherpar} shows the values
for the absorption $N_H$, the power-law photon index, the cutoff energy $E_{\rm
cut}$, and the folding energy $E_{\rm fold}$. The 2-100 keV flux is again shown
in the bottom panel for comparison. Lastly, Figure~\ref{fig:Fe} shows the iron
line parameters, the line energy, width, and normalization. In addition, we also
examined fits that included an additional Gaussian absorption feature at $E_{\rm
cut}$, the MPLCUT model of \citet{Coburn02}. An F-test showed that this
component was not significant. Lastly we tried adding a second Gaussian
absorption feature around 20 keV. We attempted two approaches, one where the
line energy was exactly twice that of the other feature and one where both line
energies were free to vary independently. F-tests showed that including a second
cyclotron feature (using either approach) did not significantly improve the fit.For a 
small number of spectra near the peak of the outburst, we also tried fitting a spectrum
including a bump near 15 keV after \citet{Klochkov07} instead of a Gaussian absorption
feature. For all six spectra we fit with this model, it was a significantly poorer
fit than the Gaussian absorption feature.

\section{Discussion}
\subsection{Evidence for a Cyclotron Feature}
The giant outburst of EXO 2030+375 began in early June 2006 and lasted
until the end of October 2006. From our spectral fitting, we consistently
detected a cyclotron feature at about 11 keV from 2006 June 22 (MJD 53908), when
pointed PCA and HEXTE observations began, until 2006 September 26 (MJD 54004).
From 2006 September 26 through 2006 October 2 (MJD 54010), we still detected the
feature in about half of the observations. After 2006 October 2, we did not
consistently detect the feature. This means that we consistently detected the
feature at luminosities greater than about $5 \times 10^{37}$ erg s$^{-1}$.

As a consistency check, we examine the expected
relationship between the magnetic field and the maximum spin up rate, $\dot \nu$, for EXO 2030+375.
\begin{eqnarray}
\dot \nu = 2.61 \times 10^{-11} {\rm Hz\ s}^{-1} k^{1/2} \alpha^{8/7} \beta^{6/7} 
\left( \frac{M_x}{1.4 M_\sun}\right)^{-3/7} I_{45}^{-1} R_6^{12/7} 
\left(\frac{d}{7.1\ {\rm kpc}}\right)^{12/7} \\
\left(\frac{E_{\rm cyc}}{11\ {\rm keV}}\right)^{8/7} \nonumber
\left(\frac{F_{\rm peak}}{3.5 \times 10^{-8}\ {\rm erg\ cm}^{-2}{\rm s}^{-1}}\right)^{6/7} \
\label{eqn:peaknudot}
\end{eqnarray}
where $k \simeq 0.5-1.0$ is a multiplier of the equation for the disk inner edge; $\alpha = (1.31 \times
0.621) = 0.890$ accounts for the redshift of the cyclotron line $1+z \simeq 1.31$ and the general relativistic
correction for the dipole field, 0.621 \citep[See][]{Wasserman83}; $\beta = L/4\pi
d^2 F$ is the beaming factor,
estimated to be 0.7-1.3, based on the range seen by eye in the pulse profiles (A more detailed study of the
beaming factor based upon detailed modeling of pulse profiles is planned for a future paper.); $M_x$ is the 
neutron star mass; $I_{45}$ is the neutron star moment of inertia in units of
10$^{45}$ g cm$^{2}$; $R_6$ is the
neutron star radius in units of $10^6$ cm; $d$ is the distance; $E_{\rm cyc}$ is the cyclotron feature 
energy; and $F_{\rm peak}$ is the peak flux. Our measured peak spin-up rate of $1.815 \times 10^{11}$ Hz 
s$^{-1}$ is consistent with Equation~\ref{eqn:peaknudot} for $k \simeq 0.5-0.9$. 

Examining Figure~\ref{fig:cycl}, we see that near the beginning of the outburst,
the cyclotron line energy was smaller for the first two observations, 10-11 keV,
then larger, about 13-14 keV. As the outburst brightened,
the line energy evolved to about 11 keV, where it remained stable for much of the
outburst. As the outburst began to fade, the cyclotron energy appeared to
decrease; however, these observations were shorter than those taken at the
beginning of the outburst, so the line parameters were less well-determined. The
line depth and width also showed some possible luminosity dependence. In 
Figure~\ref{fig:otherpar}, we see that the power-law photon index and the 
e-folding energy show evidence for a luminosity dependence. Lastly in 
Figure~\ref{fig:Fe}, only the Fe line normalization shows a luminosity dependence.
This is expected for a Fe line that is due to EXO 2030+375 and not a background
source.

\integral and \swift observations of EXO 2030+375 during the 2006 giant outburst
revealed evidence for two cyclotron features at 10 and 20 keV \citep{Klochkov07}.
The lower energy feature lies slightly below our \rxte results. We believe this
is related to problems with the GABS model in XSPEC discussed earlier. The 20 
keV feature is not detected with \rxteno. Other spectral parameters also differ between our \rxte measurements
and \integralno/\swiftno, likely due to inclusion of second line feature in the
\integralno/\swift spectrum and instrumental differences. 

During an outburst in 1998, the Be/X-ray binary XTE J1946+274 was found to have
cyclotron feature at about 35 keV \citep{Heindl01}. The cyclotron feature
appeared to have been consistently detected above $3\times 10^{-9}$ erg cm$^{-2}$
s$^{-1}$ (2-60 keV) corresponding to a luminosity of $(2-4) \times 10^{37}$
erg s$^{-1}$ for an assumed distance of 8-10 kpc \citep{Wilson03}, similar 
luminosities to where EXO 2030's feature was consistently detected. However, no 
evolution of the cyclotron feature with energy is seen in XTE J1946+274.

4U0115+63 observations with \rxte of an outburst in 1999 showed two cyclotron
features at $\sim 11$ and $\sim 22$ keV for 3-50 keV luminosities of $(5-13)
\times 10^{37}$ erg s$^{-1}$, assuming a distance of 7 kpc. As the luminosity
decreased below $\sim 5 \times 10^{37}$ erg s$^{-1}$, the second resonance
disappeared and the fundamental resonance energy gradually increased, up to
$\sim 16$ keV at $0.16 \times 10^{37}$ erg s$^{-1}$ \citep{Nakajima06}. Similar
behavior was also observed in earlier outbursts. In 1990 February, two cyclotron
features were detected at 11.3 and 22.1 keV, when the luminosity was 
$1.4 \times 10^{38}$ erg s$^{-1}$. In 1991 March, when the luminosity was 
$2.0 \times 10^{37}$ erg s$^{-1}$, the fundamental cyclotron feature was 
detected at 15.6 keV \citep{Mihara04}. In both papers, the change in the
cyclotron energy is believed to be related to a decrease in the height of the
accretion shock in response to the reduction of the accretion rate; however,
the existing models only qualitatively describe the observations. Above $\sim 7
\times 10^{37}$ erg s$^{-1}$ the accretion shock height appears to saturate and
is no longer correlated with luminosity \citep{Nakajima06}. During the rise of
its 2006 giant outburst, EXO 2030+375's cyclotron energy appears to saturate 
above about $1.2 \times 10^{38}$ erg s$^{-1}$ (2-100 keV), assuming a distance 
of 7.1 kpc. 
\subsection{Long-Term Behavior}

\begin{figure}
\plotone{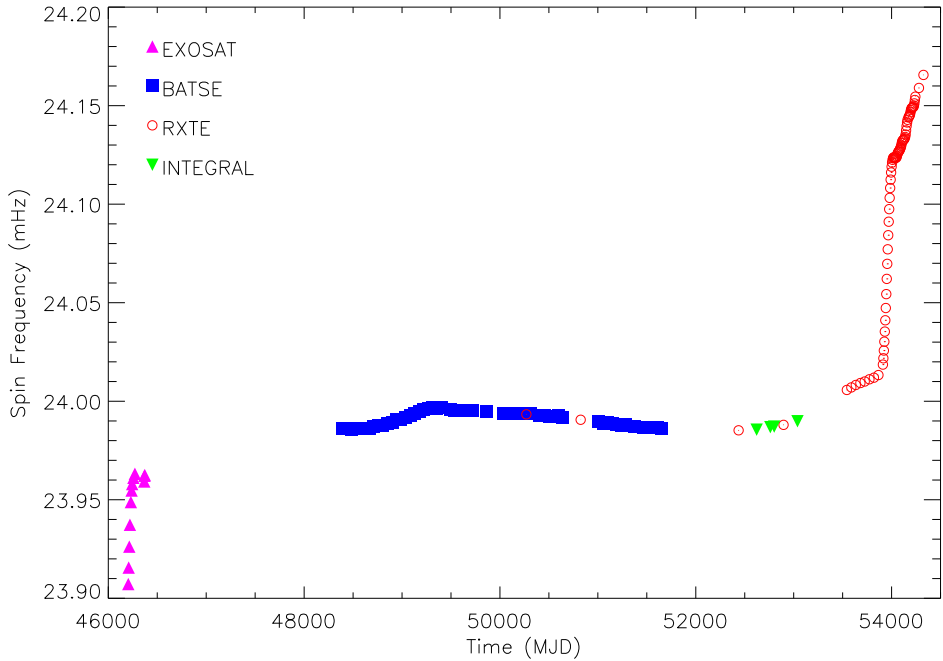}
\caption{Barycentered, orbit corrected, spin-frequency measurements for EXO 
2030+375 measured with \exosatno, BATSE, \rxte PCA, and 
\integralno.}
\label{fig:ltfreq}
\end{figure}

\begin{figure}
\plotone{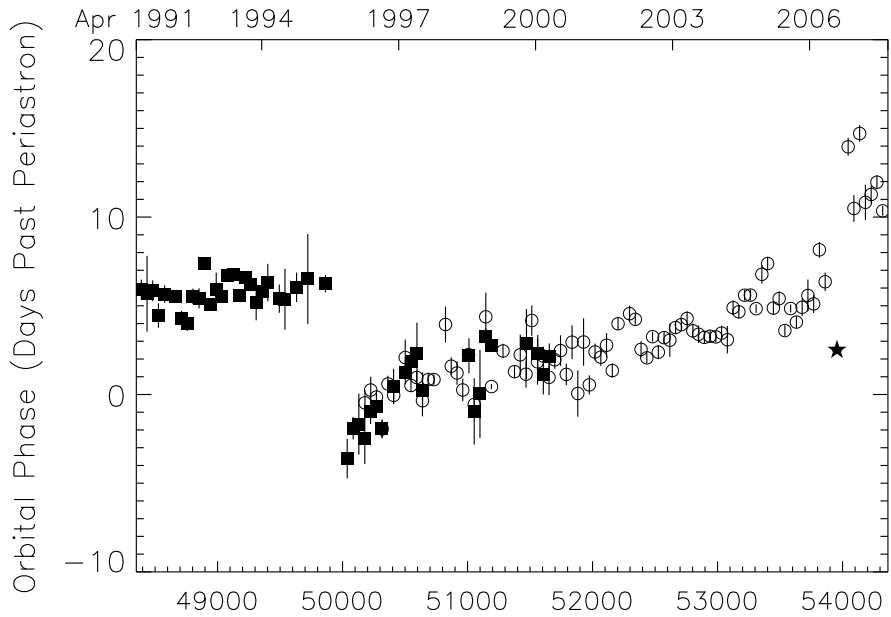}
\caption{Orbital phase (in days past perisastron) of EXO 2030+375 outburst peaks.
Normal outburst peaks measured with BATSE ({\em squares}) are taken from
\citet{Wilson02}. Normal ({\em open circles}) and giant ({\em star}) outburst peak
orbital phases were determined from Gaussian fits to \rxte ASM data. The giant
outburst peaked 1.9 orbits after the previous normal outburst peak.}
\label{fig:outbphase}
\end{figure}

\begin{figure}
\plotone{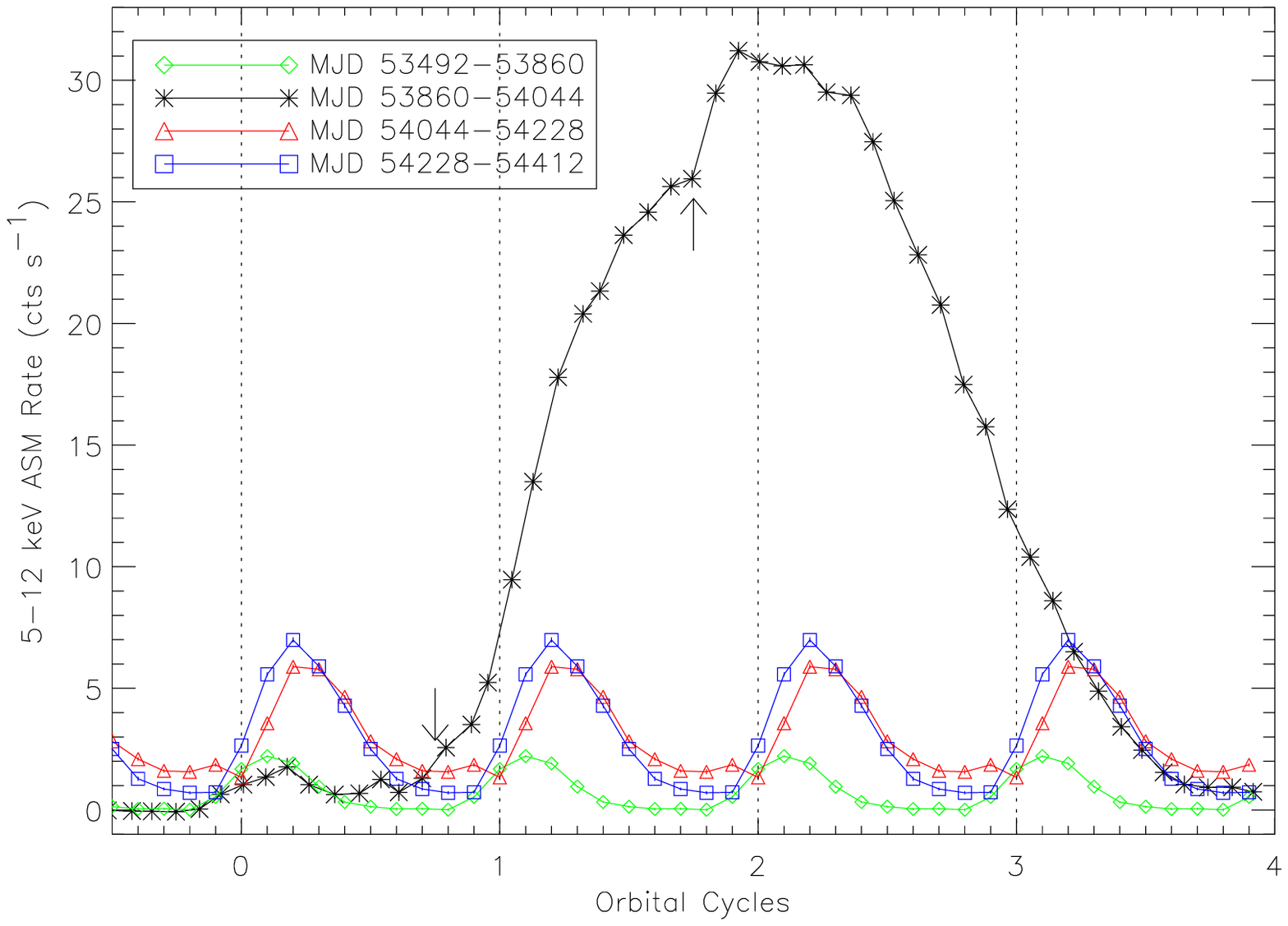}
\caption{A comparison between the giant outburst and normal outbursts before and
after it. The black line with asterisks denotes 4-day average \rxte ASM
measurements in the 5-12 keV band for the giant outburst and its precursor
normal outburst. Epoch-folded profiles, repeated 4.5 times, are shown for the
eight outbursts preceding the giant outburst (denoted green diamonds), four
outbursts immediately following the giant outburst (denoted by red triangles),
and next three outbursts following those (denoted by blue squares). Arrows
denote the approximate orbital phase of the initial rise of the giant outburst
and the approximate orbital phase of an abrupt jump in flux. Both occur at
orbital phase $\sim 0.75$.}
\label{fig:giantphase}
\end{figure}

Figure~\ref{fig:ltfreq} shows the long-term frequency history of EXO 2030+375
including measurements with {\em EXOSAT} \citep{Parmar89a}, BATSE, \integralno, and 
\rxte \citep{Wilson02,Wilson05}. Figure~\ref{fig:ltflx} shows the long-term flux 
history, including \exosat measurements from \citet{Parmar89a}. Each spike in the ASM history corresponds to a 
normal outburst. Typically, 1-4 points in the BATSE history \citep{Wilson02} correspond
to a normal outburst. The large peak is the 2006 giant outburst.

As of early September 2007, seven normal outbursts have
been observed with \rxte following the giant outburst. They continue to show 
considerable spin-up. During the first five of these outbursts, daily PCA
observations detected EXO 2030+375 pulsations throughout its orbit. For the last
two outbursts, PCA observations were taken only near the outburst peaks. 
Figure~\ref{fig:ltflx} shows that these outbursts are generally brighter than 
those just before the 2006 giant outburst. Further, the ASM observations 
continue to indicate that there is significant emission, even between the 
outbursts. Figure~\ref{fig:outbphase} shows the orbital phase of EXO 2030+375 
outbursts since 1991. The points for the BATSE outbursts are taken from 
\citet{Wilson02}. The \rxte ASM data were divided into 46-day intervals and each
interval was fit with a Gaussian to determine the outburst peak. Similarly, 
using a Gaussian fit to 120 days of ASM data, we found that the giant
outburst peaked at 2.5 days after periastron, 1.9 orbits after the previous
normal outburst. From 1991 to 1995, EXO 2030+375's outbursts consistently peaked
about 6-days after periastron. At some point between 1993 and 1995, a precessing
density perturbation developed in the Be star's disk. In late 1995, the density
perturbation interacted with the neutron star's orbit at a phase corresponding 
to 2.5 days before perisastron. The outburst peaks then quickly migrated in 
orbital phase until late 1997 when the density perturbation lost contact with 
the neutron star's orbit at a phase corresponding to 2.5 days after periastron 
\citep{Wilson02,Wilson05}. From 1997 until just before the giant outburst in 
2006, the outburst orbital phases continued to slowly migrate. Just before the 
giant outburst in 2006, the outburst peaks had finally returned to approximately six days after periastron.
After the giant outburst, the first normal outburst peaked at about 14 days
after periastron as did the third and brightest of the normal outbursts. The remaining
five normal outbursts peaked about 11 days after periastron. It is interesting to note
that both sudden shifts in outburst phase are similar in magnitude, about 
8-9 days, but opposite in sign. 

Figure~\ref{fig:giantphase} shows a comparison between the giant outburst and
the average profiles of normal outbursts before and after it. The giant outburst
and the small normal outburst that preceded it are denoted by black asterisks.
The normal outburst preceding the giant outburst was not particularly bright,
occurred at the same orbital phase as the average profile for the preceding
eight outbursts (green diamonds), but was unusual in that at about the point the
outburst should have started fading, the flux remained higher than expected and
then quickly rose into the giant outburst. Arrows on the plot denote orbital
phase 0.75, where the giant outburst appears to be starting to rise rapidly, and
where we see a rapid increase in flux one orbit later. The average profile from
the first four outbursts following the giant outburst (red triangles) shows that
those outbursts began later in orbital phase and had a higher intra-outburst 
flux than the pre-giant outbursts and the later post-giant outbursts (blue 
squares). From Figures~\ref{fig:giantphase} and \ref{fig:ltflx}, the 
intra-outburst flux is slowly fading. It appears to be a long-lived 
tail of the giant outburst. A long-lived tail of a giant outburst has not been 
observed before in Be/X-ray binaries.
 
From our orbital fits using 17 years of EXO 2030+375 data, we see marginal
evidence for apsidal precession. Our measurements are consistent with
predictions from simple calculations given below. In the supernova explosions 
that form Be X-ray binaries the neutron star will receive a kick out of the 
original orbital plane. If the orbit is misaligned with the companion's spin 
axis, precession of the orbital plane is expected. Using calculations described
in \citet{Lai95} we estimate the magnitude of the expected effect due to 
interactions between the companion's rotation period and the orbit. 
The apsidal motion and orbital plane precession change $\omega$ the observed
longitude of periastron and $i$ the orbital inclination angle. Using Equation~7
in \citet{Lai95} to calculate the apsidal rate $\dot \omega$ gives
\begin{eqnarray}
\dot \omega = 1.6 \times 10^{-3} 
\left( \frac{46.0208\ {\rm days}}{P_{\rm orb}} \right) 
\left( \frac{k}{0.01} \right)
\left( \frac{8.9 R_{\rm c} \sin i}{a_{\rm x}} \right)^2 \\
\times \left( \frac{\hat \Omega_{\rm s}}{0.5} \right) \nonumber 
\left( 1-\frac{3}{2} \sin^2 \theta \right) {\ \rm rad\ yr}^{-1}.
\end{eqnarray}
where $k \simeq 0.01$ is the apsidal motion constant for a $10 M_{\sun}$ 
main-sequence star \citep{Lai95}; $R_{\rm c} = 6 R_{\sun}$ is the assumed Be 
star radius; $\hat \Omega_{\rm s}$ is the dimensionless spin of the companion, 
assumed to be near break-up $\hat \Omega_{\rm s,max} = 0.5$; $a_{\rm x}$ is the
measured semi-major-axis; $P_{\rm orb}$ is the measured orbital period 
(see Table~\ref{tab:orb}); $i$ is the orbital inclination angle; and $\theta$ is
the angle between the orbital angular momentum and the spin angular momentum.

From our orbital fits, we measured apsidal rates ranging from 
$(2.6 \pm 0.9) \times 10^{-3}$ radians yr$^{-1}$ (Fit 3, Table~\ref{tab:orb}) 
to $(3.1 \pm 0.9) \times 10^{-3}$ radians yr$^{-1}$ (Fit 2, 
Table~\ref{tab:orb}). Our measurements are within a factor of about 2 of the 
simple calculations, suggesting that we are seeing apsidal motion in this 
system. The change in $i$ is given by Equation~8 in
\citet{Lai95} and has the same magnitude as $\dot \omega$
\begin{eqnarray}
\frac{{\rm d}i}{{\rm d}t} = 1.6 \times 10^{-3} 
\left( \frac{46.0208\ {\rm days}}{P_{\rm orb}} \right) 
\left( \frac{k}{0.01} \right)
\left( \frac{8.9 R_{\rm c} \sin i}{a_{\rm x}} \right)^2 \\
\times \left( \frac{\hat \Omega_{\rm s}}{0.5} \right) \nonumber 
\sin \theta \cos \theta \sin \Phi {\ \rm rad\ yr}^{-1}.
\end{eqnarray}
In Fits 2 and 3, we see a suggestion of a change in $i$, 
$di/dt = (0.8 \pm 0.4) \times 10^{-3} \sin^{-1} i$ radians yr$^{-1}$, also 
consistent with the calculation within a factor of about two.

\subsection{Comparison with 1985 Giant outburst}
In the discovery outburst observed with EXOSAT in 1985, the maximum observed
luminosity in the 1-20 keV band was $2 \times 10^{38}$ erg cm$^{-2}$ s$^{-1}$
\citep{Parmar89a}. The maximum 2-20 keV luminosity observed with RXTE PCA was 
$1.44 \times 10^{38}$ erg cm$^{-2}$ s$^{-1}$, about 72\% of the EXOSAT value.
Both luminosities assume a distance of 7.1 kpc \citep{Wilson02}. 
The difference cannot be explained from absorption, suggesting that the 1985 outburst
was brighter. Further evidence for this comes from comparing the peak spin-up rates
from both outbursts. The peak spin-up rate measured in the 1985 outburst was $2.4
\times 10^{-11}$ Hz s$^{-1}$, while the peak spin-up rate measured in the 2006 outburst
was $(1.815 \pm 0.006) \times 10^{-11}$ Hz s$^{-1}$, also about 75\% of the EXOSAT
value. However, despite the fainter peak, the 2-20 flux minimum reached after 
the 2006 outburst was $(4.0 \pm 0.4) \times 10^{-10}$ erg cm$^{-2}$ s$^{-1}$ 
much brighter than the 1985 upper limit of $1.3 \times 10^{-11}$ erg cm$^{-2}$ 
s$^{-1}$ (1-20 keV). The 2006 flux minimum occurred about 90 days after the peak
and similarly the 1985 upper limit was measured 98 days after the initial and 
brightest detection. Presumably, the \exosat
detections of EXO 2030+375 from 1985 October 29 - November 3 \citep{Parmar89b}
corresponded to a normal outburst. Using our best ephemeris, we find that
these observations span orbital phases of 8.5 to 13.5 days past periastron, very
similar to the orbital phases where we see normal outbursts after the 2006 giant 
outburst. The first normal outburst after the 2006 giant outburst reached a peak 2-20
keV flux of about $6.6 \times 10^{-9}$ erg cm$^{-2}$ s$^{-1}$, approximately a factor
of two brighter than the non-flaring observation on 1985 November 3 and a factor of
about 3 fainter than the brightest flares on 1985 October 30-31 \citep{Parmar89b}.
From Figure~\ref{fig:postgiantres} it is clear that considerable day to day variability
was present especially in the first outburst after the 2006 giant outburst and between outbursts.
This variability may be related to the flaring seen on 1985 October 30-31. Both
flaring episodes occurred at similar orbital phases; however, our \rxte 
observations were too short to confirm if the same flaring behavior was 
occurring.

\section{Conclusions}
EXO 2030+375 has now been observed for more than 22 years. In this time, it underwent two giant
outbursts, in 1985 and 2006, and numerous normal outbursts. Weak evidence for evolution of the binary orbit is
presented in this paper. Perhaps by the next giant outburst this can be accurately determined. 
In 2006, we observed the onset of a giant outburst of EXO 2030+375 for the first time. A normal outburst, which peaked slightly 
later than the previous ones, preceded the 2006 giant outburst. Near orbital phase 0.5 (See Figure~\ref{fig:giantphase})
the flux from EXO 2030+375 began to increase instead of declining as it usually would after a normal outburst. The flux
and spin-up rate continued to increase for more than an entire orbit. Near orbital phase 0.75, the flux abruptly 
increased, accompanied by an abrupt increase in the spin-up rate. This was followed by a relatively flat-topped 
maximum during which the flux remained within 10\% of the maximum for about 25 days, from orbital phase --0.2 to 0.35.
After that, the flux declined to a minimum more than 1.5 orbits later. The entire giant outburst spanned more than 
three pulsar orbits.

During the 2006 giant outburst, we discovered evidence for a cyclotron feature with a mean value of
$11.44\pm0.02$ keV that was consistently detected for about 90 days, at 2-100 keV luminosities above 
$5\times 10^{37}$ erg s$^{-1}$. This feature translates into a magnetic field strength of $B = 9.9 \times
10^{11} (1+z)$ G $\simeq 1.3 \times 10^{12}$ G. Predictions of the peak spin-up rate during the giant
outburst, using this magnetic field strength, are consistent with the measured peak spin-up rate during the
2006 giant outburst. \citet{Klochkov07} proposed that the EXO 2030+375 spectrum
measured with \integral and \swift could also be explained by a model containing a bump at 15 keV. However, our
\rxte observations do not appear to support that model. The cyclotron features for EXO 2030+375 and 
4U 0115+63 \citep{Nakajima06} have the lowest energies measured to date. Now we are left with the question: 
Is the lack of detections of $E_{\rm cyc} < 10$ keV a physical or observational bias ?

The two classes, normal (type I) and giant (type II), of Be/X-ray pulsar outbursts are clearly distinguishable
in the observations of EXO 2030+375. However, the evolution seen in the phasing, duration, and peak flux of
the normal outbursts, while apparently related to the Be star's disk \citep{Wilson02}, is poorly understood.
The lack of gaps in detection between the normal outbursts after the 2006 giant outburst implies
that the accretion disk present in the giant outburst persists throughout these normal outbursts.
Many have believed that the giant outbursts and normal outbursts were distinguished by the presence of an
accretion disk in the giant outburst, while the normal outbursts proceeded by wind accretion. Possibly the
normal outbursts following a giant outburst \citep[as in the ``mother duck/baby duck" complexes seen with
BATSE in][]{Bildsten97} are a distinct class of normal outbursts, with the remainder proceeding by wind accretion.
However, XTE J1946+274 \citep{Wilson03} had a series of normal outbursts that also had no detection gaps
between them, but these outbursts were not preceded by a giant outburst. \citet{Ikhsanov01} show for longer 
period sources, e.g. A0535+26, disk accretion is required for the normal
outbursts to occur. \citet{Okazaki} conclude from simulations that a transient disk forms in normal outbursts
in shorter period systems, e.g. 4U0115+63. Measurements of significant spin-up during normal outbursts of EXO
2030+375 prior to the 2006 giant outburst \citep{Wilson02, Wilson05} suggested that at least a transient 
accretion disk was also present in those outbursts.  Further, \citet{Hayasaki06} conclude from simulations that accretion disks
in Be/X-ray binaries evolve through three phases: a `developing phase' where the mass accretion rate is double-peaked, but 
dominated by direct accretion at periastron; a 'transition phase' where the mass accretion rate evolves from double-peaked to
single peaked as the approximately Keplerian disk grows with time; and finally a quasi-steady state where the mass accretion rate
has a single peak induced by a one-armed spiral wave and is on average balanced with the mass-transfer rate from the Be disk.
\citet{Camero05} reported multiple detections of an initial spike preceding the main peak in outbursts prior to the giant
outburst. This initial spike was usually smaller than the main outburst peak, suggesting that EXO 2030+375 was in the `transition
phase.' After the giant outburst, the initial spike preceding the main peak has disappeared from the normal outbursts, suggesting
that the accretion disk in EXO 2030+375 has reached the final quasi-steady state. Therefore, the long-term behavior seen in the 
normal outbursts of EXO 2030+375 appears to be product of the state of the Be disk and the accretion disk, allowing for a wide 
range of variations. 

Giant outbursts from EXO 2030+375 also vary in brightness and in orbital phase. The 2006 giant outburst peaked at 
a 2-100 keV luminosity of $2.2 \times 10^{38}$ erg s$^{-1}$. In the 2006 giant outburst, the 2-20 keV luminosity and 
the peak spin-up rate were about 72\% and 75\%, respectively, of that measured in the the 1985 outburst, suggesting 
that the 1985 outburst had a 2-100 keV peak luminosity of $(2.9-3.0) \times 10^{38}$ erg s$^{-1}$, which was 
super-Eddington unless the neutron star mass was greater than about $2.2 M_{\sun}$.  The 1985 outburst was shifted by 
0.65 in orbital phase relative to the decline of the 2006 outburst. Interestingly, the decay constants of both outbursts
were very similar. Giant outbursts in EXO 2030+375 do not appear to be locked in orbital phase, unless the onsets are.
The 2006 minimum spanned orbital phase --0.2 to --0.1, very similar to the orbital phase of the 23 August 1985 
minimum; however, the 2006 minimum was more than 30 times brighter than in 1985. Despite the differences between the 
giant outbursts, the normal outburst following the 1985 giant outburst lined up quite well in orbital phase with the 
second normal outburst following the 2006 giant outburst. Both normal outbursts also showed flaring activity.

\acknowledgements
This research has made use of data obtained from the High Energy Astrophysics
Science Archive Research Center (HEASARC), provided by NASA's Goddard Space
Flight Center (GSFC). \rxte ASM quick-look results were provided by the \rxte
ASM teams at MIT and at the GSFC SOF and GOF. We thank Morgan Dwyer, a summer
intern from Yale who performed the early timing analyses as the observations originally
came in. We also thank Evan Smith and Jean Swank for their help scheduling the
daily \rxte observations of the giant outburst and of the normal outbursts that
followed.

\end{document}